\documentclass{article} 

\usepackage{graphicx}
\graphicspath{{converted_graphics/}}
\begin{document}

\begin{center}
{\Large A Basic Thermodynamic Derivation of the Maximum Overburden Pressure Generated in Frost Heave}\vskip16pt

{\large Kenneth G. Libbrecht}\footnote{%
e-mail address: kgl@caltech.edu\newline
}\vskip4pt

{Department of Physics, California Institute of Technology}\vskip-1pt

{Pasadena, California 91125}\vskip-1pt

\vskip18pt

\hrule \vskip1pt \hrule
\vskip 14pt
\end{center}

\noindent \noindent \textbf{ABSTRACT}

I describe a simple \textquotedblleft heat-engine\textquotedblright\
derivation of the maximum overburden pressure that can be generated in frost
heave. The method stems from the fact that useful work can, in principle, be
extracted from the forces generated by an advancing solidification front via
the frost heave mechanism. Using an idealized frost heave \textquotedblleft
engine,\textquotedblright\ together with the maximum thermodynamic
efficiency of any heat engine, one can derive the maximum overburden
pressure. A similar argument can also produce the maximum \textquotedblleft
thermodynamic buoyancy\textquotedblright\ force on a foreign object within a
solid surrounded by a premelted layer.

\section{A Frost Heave Engine}

Frost heave is a common environmental process in which the freezing of water
into ice can produce forces large enough to seriously damage roads and
bridges \cite{dash1}. Contrary to common belief, frost heave is not the
result of the simple expansion that takes place when water freezes. Rather,
it results when water migrates through a porous soil, forming ice
\textquotedblleft lenses\textquotedblright\ within the soil. This migration
tends to extract water from the pores in the soil, so the final structure
(containing soil plus ice lenses) has a higher volume than the initially
saturated soil. Frost heave has also been observed with other materials
freezing in porous media  \cite{dash2, mizusaki}.

The expansion that results from frost heave is known to produce a maximum
overburden pressure $P_{\max }=q_{m}\rho \Delta T/T_{m}$, where $q_{m}$ is
the latent heat of fusion per unit mass, $\rho $ is the ice density, $T_{m}$
is the bulk melting temperature, and $\Delta T=T_{m}-T_{1},$ where $T_{1}$
is the temperature at which the ice freezes. This is a well-known result
that is typically derived from the modified Clausius-Clapeyron equation
(e.g. \cite{dash1, mizusaki}), in which solidification gains the free-energy 
$\Delta S\Delta T,$ where $\Delta S\approx q_{m}\rho AL/T_{m}$ is the
entropy change upon freezing.

We can produce a very simple derivation of this same result by constructing
the frost heave \textquotedblleft engine\textquotedblright\ shown in Figure
1, in which mechanical work is extracted from heat flow. Referring to Figure
1(a), we have a porous piston separating ice from water in an ideal
(frictionless), insulating cylinder of length $L$ and area $A$. As the
solidification front pushes the piston down the length of the cylinder, a
heat $\delta Q\approx q_{m}\rho AL$ must be removed from the left end at
temperature $T_{1}$. If the piston travels against an external force $F,$
then a total work $\delta W=FL$ is extracted. After this \textquotedblleft
power stroke,\textquotedblright\ we complete the engine cycle by warming the
ice to $T_{m}$, melting the ice, lowering the water temperature back to $%
T_{1}$, and flipping cylinder around. We ignore the small heat needed to
change the temperature from $T_{1}$ to $T_{m}$ and back, so the dominant
heat input is again $\delta Q,$ which is applied at a temperature $T_{m}$.
Thus, after one cycle of the engine, there has been a net heat flow of $%
\delta Q\approx q_{m}\rho AL$ from $T_{m}$ to $T_{1},$ from which we
extracted a work $FL.$ Since the maximum efficiency of any heat engine is $%
\varepsilon =\Delta T/T,$ we must have $F_{\max }=q_{m}\rho A\Delta T/T_{m},$
where $\Delta T=T_{m}-T_{1},$ which is valid to first order in $\Delta T.$
This gives the maximum overburden pressure $P_{\max }=F_{\max }/A=q_{m}\rho
\Delta T/T_{m},$ equal to the expression above. 

Assuming ice does not enter the porous piston, the motion of the
solidification front can be halted by applying a force $F_{\max }$ to the
piston. If the motion of the piston is quasi-static, and the only heat flow
through the piston is via liquid flow, then a near-maximum efficiency can be
attained, and this is seen in frost heave experiments using both water and
helium \cite{dash1, mizusaki}.

We can apply the same reasoning to the situation shown in Figure 1(b), for
which we now have ice on either side of the piston, with $T_{2}>T_{1}$. In
this case, the motion of the piston does not change the total amount of ice,
except for a negligible change in the thickness of the premelted layer. As
melting and resolidification pushes the piston across the cylinder, a heat $%
\delta Q\approx q_{m}\rho AL$ is added to one side of the piston to melt the
ice at temperature $T_{2},$ and the same heat is extracted at $T_{1}$ to
refreeze the ice (again ignoring the small heat required to change the
temperature of the water). As before, there is a flow of heat $\delta
Q\approx q_{m}\rho AL$ from $T_{2}$ to $T_{1}$, from which we extract the
useful work $\delta W=FL.$ At maximum efficiency, this yields $F_{\max
}/A=q_{m}\rho \Delta T/T$, where $\Delta T=T_{2}-T_{1}.$ 

The idealized piston in Figure 1(b) is functionally equivalent to a foreign
(nonporous) particle surrounded by a premelted layer, in which case the
fluid flow is around the particle rather than through a porous piston. If
the particle has a characteristic size $R,$ then we can take $A\approx R^{2}$
and $\Delta T=R\nabla T$, where $\nabla T$ is the temperature gradient at
the object. This gives $F_{\max }=m_{s}q_{m}\nabla T/T_{m},$ where $m_{s}$
is the mass of the surrounding solid displaced by the object, which is the
result reported in \cite{rempel}.

\medskip

\begin{figure}[h] 
  \centering
  \includegraphics[bb=0 0 696 559,width=4.5in,keepaspectratio]{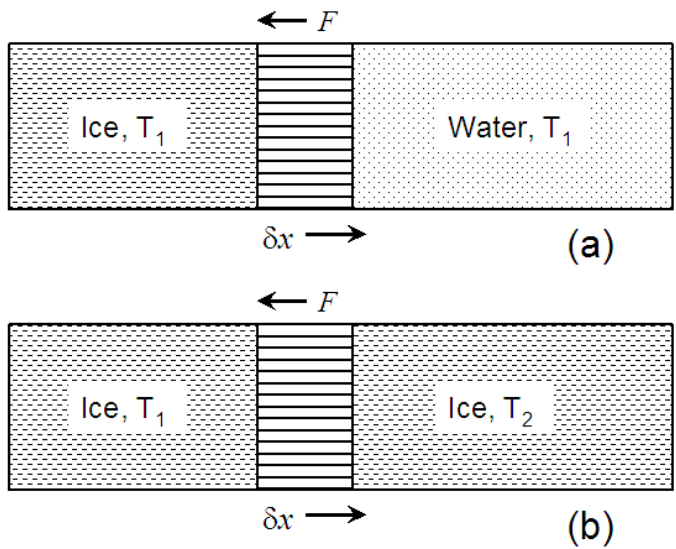}
  \caption{(a) An idealized frost-heave
experiment, in which ice at a temperature $T_{1}<T_{m}$ is separated from
water at the same temperature by a porous plug, all within a frictionless,
insulating cylinder. As water permeates the plug and solidifies, the
solidification front pushes the plug to the right, against an opposing force 
$F.$ (b) A similar situation to (a), but with ice at different temperatures
on both sides of the cylinder. Here we assume there exists a thin premelted
layer on all surfaces within the plug, which allows fluid flow from one side
to the other.}
  \label{fig:figure1a}
\end{figure}


\medskip

\section{References}
\begin{thebibliography}{9}
\bibitem{dash1} J. G. Dash, J. Fu, and J. S. Wettlaufer, \textquotedblleft
The Premelting of Ice and its Environmental Consequences,\textquotedblright\
Rep. Prog. Phys. 58, 115-167 (1995).

\bibitem{dash2} J. G. Dash, \textquotedblleft Frost Heave in Helium and
Other Substances,\textquotedblright\ J. Low Temp. Phys. 89, 277 (1992).

\bibitem{mizusaki} T. Mizusaki and M. Hiroi, \textquotedblleft Frost Heave
in Helium,\textquotedblright\ Physica B 210, 403 (1995).

\bibitem{rempel} A. W. Rempel, J. S. Wettlaufer, and M. G. Worster,
\textquotedblleft Interfacial Premelting and the Thermomolecular Force:
Thermodynamic Buoyancy,\textquotedblright\ Phys. Rev. Lett. 87, 088501
(2001).

\end{thebibliography}
\end{document}